\documentclass[12pt,a4]{article}
\usepackage{epsfig}
\begin{document}
\begin{center}
{\bf\large Forward observables at\\ RHIC, the Tevatron run II and the
LHC\footnote{presented at the Second International "Cetraro" Workshop \& NATO
Advanced Research Workshop "Diffraction 2002", Alushta, Crimea, Ukraine, August
31 - September 6, 2002.}}\\
{}~\\

{J.~R. Cudell\footnote{Institut de Physique,
B\^at. B5, Universit\'e de Li\`ege, Sart Tilman, B4000 Li\`ege, Belgium}, V.~V.
Ezhela\footnote{COMPAS group, IHEP, Protvino, Russia}, P. Gauron\footnote{LPNHE
(Unit\'e de Recherche des Universit\'es
Paris 6 et Paris 7, Associ\'ee au CNRS)-Theory Group, Universit\'e
Pierre et Marie Curie, Tour 12 E3, 4 Place Jussieu, 75252
Paris Cedex 05, France}, K. Kang\footnote{Physics Department, Brown University,
Providence,
RI, U.S.A.}, Yu.~V. Kuyanov$^2$,  S.~B. Lugovsky$^2$, E.
Martynov$^{1,}$\footnote{On leave from Bogolyubov Institute for Theoretical
Physics, 03143
Kiev, Ukraine}, B. Nicolescu$^3$, E.~A. Razuvaev$^2$, N.~P.~Tkachenko$^2$}\\
{}~\\

{COMPETE\footnote{COmputerised Models, Parameter Evaluation for Theory and
Experiment.
} collaboration}
\end{center}
\begin{abstract}
We present predictions on the total cross sections
and on the ratio of the real part to the imaginary part of the
elastic amplitude
($\rho$ parameter) for present and future $pp$ and $\bar p p$
colliders, and on total cross sections
for $\gamma p \to$ hadrons at cosmic-ray energies and
for $\gamma\gamma\to$ hadrons up to $\sqrt{s}=1$ TeV.
These predictions are based on a
study of many possible analytic parametrisations
and invoke the current hadronic dataset
at $t=0$.
The uncertainties on total cross sections, including the systematic theoretical
 errors,
reach $1\%$ at RHIC, $3\%$ at the Tevatron, and $10\%$ at the LHC,
whereas
those on the $\rho$ parameter are respectively $10\%$, $17\%$, and $26\%$.
\end{abstract}
This report is based on ref. \cite{PRL}, which constitutes the conclusion
of an exhaustive study \cite{Cudell2002} of analytic
parametrisations of soft forward data at $t=0$.
As explained in V.~V. Ezhela's contribution to these proceedings,
this study has three main
purposes. First of all, it helps maintain the dataset of cross sections
and $\rho$ parameters available to the community. Secondly, it enables
us to decide which models are the best, and in which region of $s$. Finally,
and this will be the main object of this report, it enables us to make
predictions based on a multitude of models, and on all available data.

The dataset of this study includes all measured total cross
sections and ratios of the real part to the imaginary part of the elastic
amplitude
(\( \rho  \) parameter) for the scattering of \( pp, \) \( \overline{p}p \),
\( \pi ^{\pm }p \), \( K^{\pm }p \), and total cross sections for \( \gamma p
\),
\( \gamma \gamma  \) and \( \Sigma ^{-}p \). Compared with the 2002
Review of Particle Properties dataset \cite{PDB}, it includes the
latest ZEUS points\cite{ZEUS} on total cross sections, as well as
cosmic ray measurements \cite{cosmic}. The number of points of
each sub-sample of the dataset is given in Table \ref{table1}.
\begin{table}
\begin{center}
\caption{Summary of the quality of the fits for different scenarios
considered in this report, for $\sqrt{s}\geq 5$ GeV:
DB02Z -- the 2002 Review of Particle Properties database with new
ZEUS data, DB02Z-CDF -- with
the CDF point removed;
DB02Z-E710/E811 -- with E710/E811 points
removed. The first line gives the overall $\chi^2/dof$ for the
global fits, the other lines give the $\chi^2/nop$ for data sub-samples,
the last line gives in each case the parameter controlling the asymptotic
form of cross sections.}
{}~\\~\\
\begin{tabular}{lcccc}
\hline
 & Number &DB02Z &DB02Z &DB02Z \\
Sample & of points & &$-$CDF &$-$E710/E811\\
\hline
total && 0.965 & 0.964 & 0.951 \\
\hline
\multicolumn{4}{c}{total cross sections}\\
\hline
$p p$            &111     & 0.84 & 0.90 & 0.90 \\
$\overline{p} p$ &57$-$59& 1.15 & 1.12 & 1.05 \\
$\pi^+ p$        &50      & 0.71 & 0.71 & 0.71 \\
$\pi^- p$        &95      & 0.96 & 0.96 & 0.96 \\
$K^+ p$          &40      & 0.71 & 0.71 & 0.71 \\
$K^- p$          &63      & 0.62 & 0.62 & 0.61 \\
$\Sigma^- p$     &9       & 0.38 & 0.38 & 0.38 \\
$\gamma p$       &37      & 0.58 & 0.58 & 0.58 \\
$\gamma \gamma$  &38      & 0.64 & 0.64 & 0.63 \\
\hline
\multicolumn{5}{c}{elastic forward Re/Im}\\
\hline
$p p$            &64      & 1.83 & 1.83 & 1.80 \\
$\overline{p} p$ &11      & 0.52 & 0.52 & 0.53 \\
$\pi^+ p$        &8       & 1.50 & 1.52 & 1.46 \\
$\pi^- p$        &30      & 1.10 & 1.09 & 1.14 \\
$K^+ p$          &10      & 1.07 & 1.10 & 0.98 \\
$K^- p$          &8       & 0.99 & 1.00 & 0.96 \\
\hline
\multicolumn{5}{c}{ values of the parameter B }\\
\hline
& &0.307(10)&0.301(10)&0.327(10)\\
\hline
\end{tabular}
\label{table1}
\end{center}
 \end{table}

The base of models is made of 256 different analytic parametrisations.
We can summarize their general form by quoting
the form of total cross sections, from which the $\rho$ parameter is obtained
via derivative dispersion relations. The ingredients are
the contribution $M^{ab}$ of the highest
meson trajectories ($\rho$,
$\omega$, $a$ and $f$) and the rising term $H^{ab}$ for the pomeron.
\begin{equation}
\sigma_{tot}^{ab}=(M^{ab}+H^{ab})/s
\end{equation}
The first term is parametrised via Regge theory, and we allow the lower
trajectories to be partially non-degenerate, {\it i.e.} we allow one intercept
for the $C=+1$ trajectories, and another one for the $C=-1$ \cite{CKK}.
A further lifting of the degeneracy is certainly possible, but does not
seem to modify significantly the results \cite{Martynov2002}. Hence we
use
\begin{equation}
M^{ab}= Y^{ab}_{+} \left({s\over s_0}\right)^{\alpha _{+}}
\pm
Y^{ab}_{-} \left({s\over s_0}\right)^{\alpha _{-}}
\label{lower}
\end{equation}
with $s_0=1\ {\rm GeV}^2$.
The contribution of these trajectories is represented by RR.
As for the pomeron term, we choose a combination of the following
possibilities:
\begin{eqnarray}
{\textrm{H}}^{ab}&=&X^{ab}\left({s\over s_0}\right)^{\alpha _{\wp }}
+s P^{ab}
\label{pom1}\\
{\textrm{H}}^{ab}&=&s\left[B^{ab}\ln \left({s\over s_0 }\right)
+P^{ab}\right]
\label{pom2}\\
{\textrm{H}}^{ab}&=&s\left[B^{ab}\ln^{2}\left({s\over s_1}\right)
+P^{ab}\right]\label{pom3}
\end{eqnarray}
with $s_0=1\ {\rm GeV}^2$ and $s_1$ to be determined by the fit.
The contribution of these terms is marked PE, PL and PL2 respectively.
Note that the pole structure of the pomeron cannot be directly obtained from
these forms, as multiple poles at $J=1$ produce constant terms which mimic
simple poles at $t=0$.
Furthermore, we have considered several possible constraints on the parameters
of Eqs.~(\ref{lower}-\ref{pom3}):
\begin{itemize}
\item degeneracy of the reggeon trajectories \( \alpha _{+}=\alpha _{-} \),
noted (RR)$_d$;
\item universality of rising terms ($B^{ab}$ independent of the hadrons), noted
L2$_u$, L$_u$ and E$_u$ \cite{universal};
\item factorization for the residues in the case of the \( \gamma \gamma  \)
and \( \gamma p \) cross sections. If not otherwise indicated by the subscript
$nf$, we
impose \( H_{\gamma \gamma }=\delta H_{\gamma p}=\delta ^{2}H_{pp} \);
\item quark counting rules \cite{qc} to predict the $\Sigma p$
cross section from
$pp$, $Kp$ and $\pi p$, indicated by the subscript $qc$;
\item Johnson-Treiman-Freund \cite{jtf}
relation for the cross section differences, noted $R_c$.
\end{itemize}
All possible variations of Eqs.~(\ref{lower}-\ref{pom3}), using the above
constraints, amount to 256 variants.

These variants are then fitted to the database, allowing for the
minimum c.m. energy $\sqrt{s_{min}}$ of the fit to vary between
3 and 10 GeV.
For $\sqrt{s}\geq 9$ GeV,
33 variants have an overall $\chi^2/d.o.f.\leq 1.0$ if one fits
only to total cross sections, whereas 21 obeyed this criterion when
one includes the $\rho$ parameters in the data to be fitted to.
One can try to lower the minimum energy of the fit, and one finds
that for 11 models one can extend the minimum energy of the cross
section fit to 4 GeV, and that of the combined fit of $\sigma_{tot}$
and $\rho$ to 5 GeV.
Several
parametrisations based on triple poles (RRPL2), double poles (RRPL)
or simple poles (RRPE) are kept. The only notable candidate which
seems to be ruled out is the popular simple-pole model (RRE)
\cite{DoLa}. Its predictions for $pp$ and $\bar p p$
nevertheless fall within our errors\footnote{This conclusion could
be affected by a re-calculation of the $\rho$ parameter from
{\it integral} dispersion relations -- see O.~V. Selyugin's contribution to
these proceedings.
}.

After this selection is made, the remaining models are ranked. We measure some
characteristics of the fits, namely: the number of parameters, the
confidence level in the considered region, the size of the region
where the model achieves a $\chi^2/dof \leq 1$ and the value of
that $\chi^2/dof$, the stability of the parameters
 when the minimum c.m. energy is changed, their stability
with respect to the inclusion of the $\rho$ data, the uniformity of the
$\chi^2/dof$ for different processes and quantities,
and finally the quality
of the correlation matrix. All these features are important, and
we have managed to measure them, introducing new statistical
indicators \cite{Cudell2002,Cudell}.
The ideal fit would be the one with the least number of parameters,
the biggest region of applicability, the best $\chi^2$, etc. Unfortunately,
a single fit does not concentrate all these virtues.
As the new indicators do not have (yet) a probabilistic interpretation
and as all the parametrisations which fit are {\it a priori} acceptable,
we choose the ``best'' model through a ranking procedure: for each
feature, the models are ordered according to how well they perform.
One then sums the position of each model for each indicator, and
the model with least points is preferred. The advantage of this
method, besides the fact that it automatically looks at many qualities
of each fit, is that the best model is decided on the basis of
automatic criteria, which do not depend on our own prejudice.

Following that procedure, the triple-pole parametrisation
RRP$_{nf}$L2$_u$  \cite{universal,Gauron:2000ri} gives the most satisfactory
description of the data.
This parameterization has a universal (u)
$B\ln^2(s/s_0)$ term, a non-factorizing (nf) constant term and non-degenerate
lower trajectories.

\begin{figure}
\begin{center}
\epsfig{file=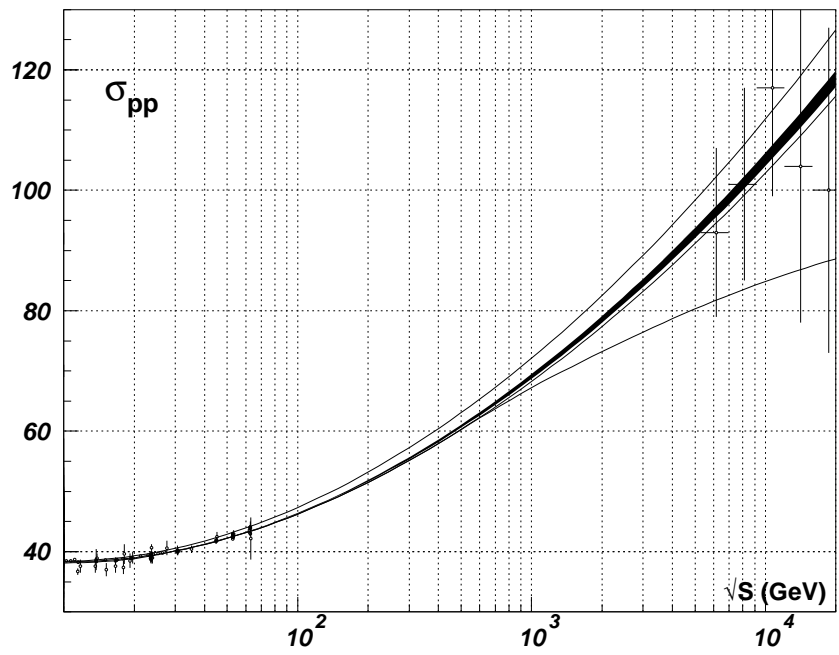,width=11cm}
\caption{Predictions for total cross sections.
}
\label{fig1}
\end{center}
\end{figure}

\begin{figure}
\begin{center}
\epsfig{file=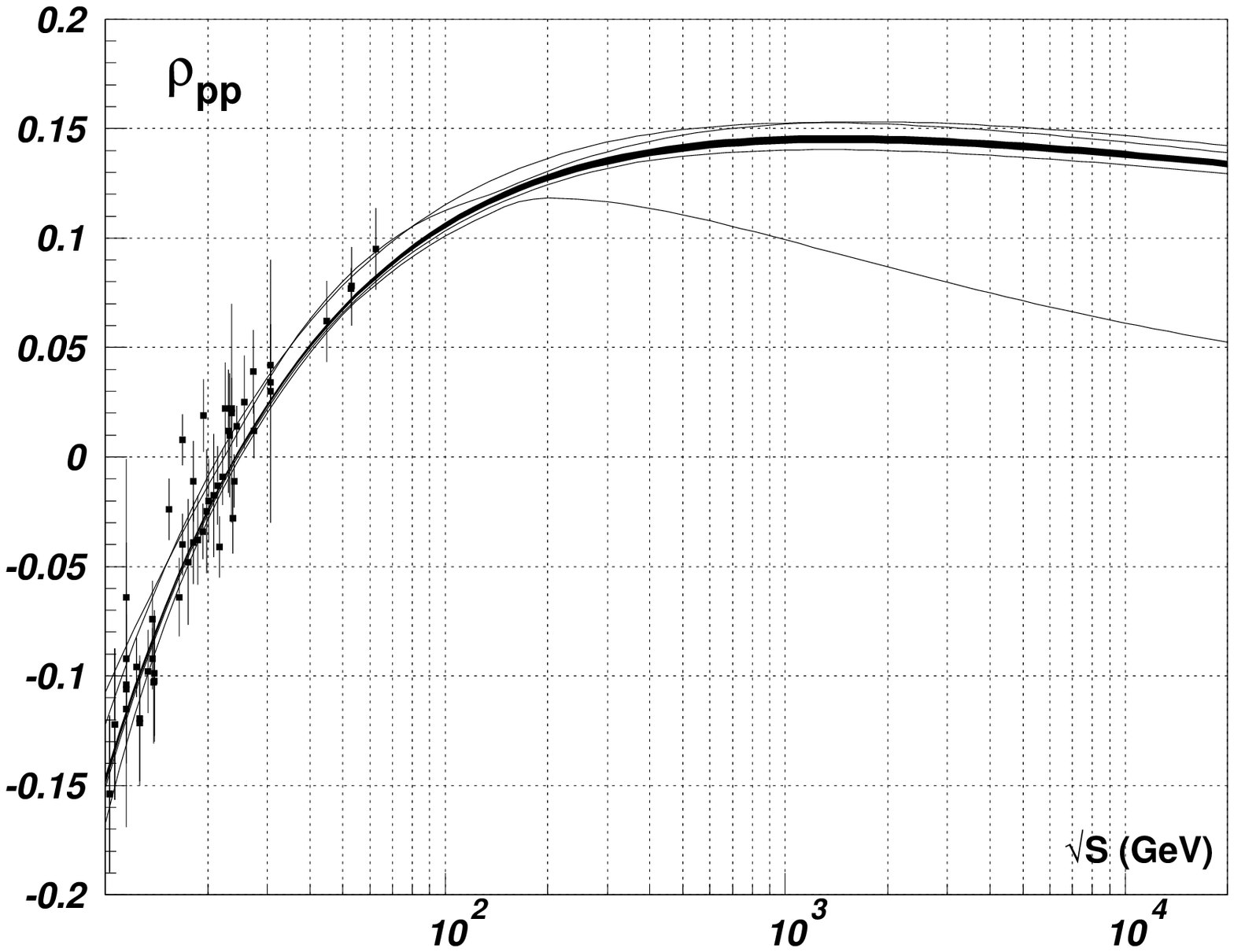,width=11cm}
\caption{Predictions for the $\rho$ parameter.
}
\label{fig2}
\end{center}
\end{figure}

We are now in a position to evaluate several quantities of interest
for future measurements. First of all, our best parametrisation
can of course be used to predict $\sigma_{tot}$ and $\rho$, with
their statistical errors. We choose for this the parameters determined
for a minimum c.m. energy $\sqrt{s_{min}}=5$ GeV. For $pp$ and
$\bar p p$, the central value of this fit gives
\begin{equation}
\sigma_{tot}^{\bar p p, pp}=43.\ s^{-0.46}\pm 33.\ s^{-0.545}+
35.5+0.307 \ln^2\left({s\over 29.}\right)
\end{equation}
with all coefficients in mb and $s$ in GeV$^2$.
We assign errors by using the full error matrix $E_{ij}$
from the fit, and define
\begin{equation}
\Delta Q=\sum_{ij} E_{ij} {\partial Q\over\partial x_i \partial x_j}
\end{equation}
with $Q=\sigma_{tot}$ or $\rho$ and $x_i$ the parameters of the model.
Our predictions are given in Table \ref{table2}
and the corresponding 1 $\sigma$ region is shown as a dark band
in Figs. \ref{fig1} and \ref{fig2}.
\begin{table}
\begin{center}
\caption{Predictions for $\sigma_{tot}$ and $\rho$, for $\bar pp$ (at
$\sqrt{s}=1960$ GeV) and for $pp$ (all other energies). The central values
and statistical errors correspond to the preferred model RRP$_{nf}$L2$_u$,
fitted for $\sqrt{s_{min}}=5$ GeV. The first systematic errors come
from the consideration of two choices between
CDF and E-710/E-811 ${\overline p} p$ data in the simultaneous global fits.
The second systematic error corresponds to the consideration of the 21
parametrisations compatible with existing data.}
{}~\\~\\
\begin{tabular}{ccc}
\( \sqrt{s} \) (GeV)& \( \sigma \) (mb)& \( \rho \)\\
\hline
  100 &$ 46.37\pm 0.06 \begin{array}{c} +0.11 \\ -0.03\end{array}
\begin{array}{c} +0.31\\ -0.06\end{array}$&$0.1058\pm 0.0012
\begin{array}{c} +0.0028\\ -0.0009\end{array}\begin{array}{c}
+0.0024\\ -0.0019\end{array}$\\
  200 &$ 51.76\pm 0.12 \begin{array}{c} +0.27 \\ -0.08\end{array}
\begin{array}{c} +0.43\\ -0.15\end{array}$&$0.1275\pm 0.0015
\begin{array}{c} +0.0035\\ -0.0011\end{array}\begin{array}{c}
+0.0000\\ -0.0023\end{array}$\\
  300 &$ 55.50\pm 0.17\begin{array}{c} +0.39\\ -0.12\end{array}
\begin{array}{c} +0.39\\ -0.20\end{array}$&$0.1352\pm 0.0016
\begin{array}{c} +0.0038\\ -0.0012\end{array}\begin{array}{c}
+0.0000\\ -0.0059\end{array}$\\
  400 &$ 58.41\pm 0.21\begin{array}{c} +0.49\\ -0.16\end{array}
\begin{array}{c} +0.28\\ -0.23\end{array}$&$0.1391\pm 0.0017
\begin{array}{c} +0.0039\\ -0.0013\end{array}\begin{array}{c}
+0.0002\\ -0.0087\end{array}$\\
  500 &$ 60.82\pm 0.25\begin{array}{c} +0.58\\ -0.19\end{array}
\begin{array}{c} +0.15\\ -0.25\end{array}$&$0.1413\pm 0.0017
\begin{array}{c} +0.0040\\ -0.0013\end{array}\begin{array}{c}
+0.0006\\ -0.0109\end{array}$\\
  600 &$ 62.87\pm 0.28\begin{array}{c} +0.66\\ -0.21\end{array}
\begin{array}{c} +0.03\\ -0.26\end{array}$&$0.1427\pm 0.0018
\begin{array}{c} +0.0040\\ -0.0013\end{array}\begin{array}{c}
+0.0008\\ -0.0125\end{array}$\\
 1960 &$ 78.26\pm 0.55\begin{array}{c} +1.30\\ -0.42\end{array}
\begin{array}{c} +0.08\\ -1.95\end{array}$&$0.1450\pm 0.0018
\begin{array}{c} +0.0038\\ -0.0013\end{array}\begin{array}{c}
+0.0022\\ -0.0226\end{array}$\\
10000 &$ 105.1\pm  1.1\begin{array}{c} +2.6 \\ -0.82\end{array}
\begin{array}{c} +0.60\\ -8.30\end{array}$&$0.1382\pm 0.0016
\begin{array}{c} +0.0032\\ -0.0011\end{array}\begin{array}{c}
+0.0028\\ -0.0324\end{array}$\\
12000 &$ 108.5\pm  1.2\begin{array}{c} +2.7 \\ -0.87\end{array}
\begin{array}{c} +0.70\\ -9.20\end{array}$&$0.1371\pm 0.0015
\begin{array}{c} +0.0031\\ -0.0011\end{array}\begin{array}{c}
+0.0030\\ -0.0332\end{array}$\\
14000 &$ 111.5\pm  1.2\begin{array}{c} +2.9 \\ -0.92\end{array}
\begin{array}{c} +0.80\\ -10.2\end{array}$&$0.1361\pm 0.0015
\begin{array}{c} +0.0030\\ -0.0011\end{array}\begin{array}{c}
+0.0030\\ -0.0337\end{array}$\\
\end{tabular}
\label{table2}
\end{center}
 \end{table}

One can now concentrate on the evaluation of systematic errors.
The first source of these is the presence of contradictory data
points in the database. One can see from Table \ref{table1} that
$\sigma_{\bar p p}$, $\rho_{pp}$, $\rho_{\pi p}$ and $\rho_{K^+p}$
are not well fitted to. For the $\rho$ parameters, this can partially
be attributed to contradictions in the data, and partially to our use
of derivative dispersion relations. For $\sigma_{\bar p p}$, this comes
entirely from problems with the data. Of these, the most notable one
is the disagreement at the Tevatron between the measurements of
CDF and those of E710/E811. We show in Table \ref{table1} the effect
of the removal of the CDF or of the E710/E811 measurements. First
of all, one can see that the coefficient of the $\log^2 s$ changes
by more than $1 \sigma$, and hence the predictions for the LHC
are muddled by this discrepancy. We also see that our preferred
parametrisation favours the CDF measurement, as its global $\chi^2/dof$
goes down when the E710/E811 point is removed. It must be noted that a
similar situation exists for all $\log^2$ and simple pole
parametrisations\footnote{The original Donnachie-Landshoff parametrisation
did predict a low cross section at the Tevatron, but this was due
to the use of a non-conventional (and non-probabilistic) $\chi^2$.
Using the statistical definition of $\chi^2$ leads to a rejection
of the E710/E711 point.}. On the other hand, the dipole RRPL ($\log s$)
parametrisations do prefer the lower value. The only thing we can do
at present is indicate what shift the adoption of one point or the others
would cause on our central value. This is given in Table \ref{table2}
as a systematic experimental error, corresponding to the shift in the upper and
lower 1$\sigma$ allowed values\footnote{This definition differs
from that of \cite{PRL} by about 1 statistical error}, and shown
as the two curves closest to the central band
in Figs. \ref{fig1} and \ref{fig2}.
 \begin{table}
\begin{center}
\caption{Predictions for $\sigma_{tot}$ for $\gamma p \to hadrons$ and for
$\gamma\gamma\to hadrons$ for
cosmic ray energies. We quote the central values, the statistical errors and
the experimental systematic errors, defined as in Table \ref{table2}.}
{}~\\~\\
\begin{tabular}{cccc}
\( p_{lab}^{\gamma} \) (GeV)& \( \sigma_{\gamma p} \) (mb)&
\( \sqrt{s} \) (GeV)& \( \sigma_{\gamma\gamma} \) ($\mu$ b)\\
\hline
$0.5\cdot10^6$&\( 0.24\pm 0.01\begin{array}{c}+0.00\\ -0.11\end{array}\)&
200&\( 0.55\pm 0.03 \begin{array}{c}+0.00\\-0.29 \end{array}\)\\

$1.0\cdot10^6$&\( 0.26\pm 0.01 \begin{array}{c}+0.00\\ -0.11\end{array}\)&
300&\( 0.61\pm 0.04 \begin{array}{c}+0.00\\-0.29\end{array} \)\\

$1.0\cdot10^7$&\( 0.33\pm 0.02 \begin{array}{c}+0.00\\-0.11\end{array}\)&
400&\( 0.66\pm 0.04 \begin{array}{c}+0.00\\-0.29 \end{array}\)\\

$1.0\cdot10^8$&\( 0.42\pm 0.02\begin{array}{c}+0.00\\ -0.11\end{array}\)&
500&\( 0.70\pm 0.05 \begin{array}{c}+0.00\\-0.29 \end{array}\)\\

$1.0\cdot10^9$&\( 0.52\pm 0.03\begin{array}{c}+0.00\\ -0.11\end{array}\)&
1000&\( 0.84\pm 0.07 \begin{array}{c}+0.00\\-0.29 \end{array}\)\\
\hline
\end{tabular}
\label{table3}
\end{center}
\end{table}

We can now, and maybe for the first time, give a reasonable estimate
of the theoretical error. The idea is to choose a less constraining
minimum energy for the fits (we take 9 GeV), and to consider the
results of the 21 models that succeed in reproducing both $\sigma_{tot}$
and $\rho$.  This gives us 21 predictions with error bars. We can
then define the theoretical systematic error by taking the distance
between the highest (resp. lowest) values in the
$1\sigma$ intervals with the 1 $\sigma$ central region. We give
the resulting numbers as a third error in Table \ref{table2}, and
as the outer curves of Figs. \ref{fig1} and \ref{fig2}.

Note that the systematic errors
cannot be added in quadrature, and that the theoretical
systematic error is an absolute
shift from model to model, and does not have any
probabilistic interpretation.

One can see that the errors on total cross sections are of the order
of 1\% at RHIC, $3\%$ at the Tevatron and as large as $10\%$ at the LHC.
At RHIC, the systematic errors dues to the Tevatron discrepancy
and those due to theory are of comparable order. The value of the
cross section is constrained by the  $\bar p p$ data, and by the fact
that we allow only one $C=-1$ contribution, which is well constrained by
the overall fit. Very precise RHIC measurements, at the level of one in
a thousand
 could shed light
on the Tevatron discrepancy, and discriminate between models.
Of course, the extrapolation to LHC energies
presents the largest uncertainty and is dominated by systematic theoretical
errors, with the double pole models (RRPL) giving
a cross section significantly lower than the triple poles or the simple poles.
A determination of the cross section at the 5\% level could rule
out one half of the models or the other.

The errors on the $\rho$ parameter are much larger, reaching 10$\%$ at RHIC,
17$\%$ at the Tevatron and 26$\%$ at the LHC. This is due to the fact that
experimental errors are bigger, hence less constraining, but this also stems
from the incompatibility of some low-energy determinations of $\rho$
\cite{Cudell2002}, and from our use of derivative dispersion relations.
Although integral dispersion relations have the potential to reduce
the $\chi^2/dof$, they have the inconvenient of introducing extra
parameters (because they necessitate subtractions). Hence it is unlikely
that a different theoretical treatment can reduce the errors. On the other
hand, a re-analysis of some of the data could be envisaged. It should
involve a combination of the information on
total cross section with that on elastic hadronic cross sections,
on electromagnetic form factors and on Regge trajectories (see
V.~V. Ezhela's contribution to these proceedings).

Finally, we can use the same approach to predict cross sections for
cosmic photon studies. We show the results in Table \ref{table3},
where we have given only the experimental systematic error\footnote{
note that the corresponding table in ref. \cite{PRL} has a typo in
the error bars which are systematically a factor 10 too small.}.

To conclude, we believe that we have pushed the database technology
to the point where it can make predictions, and decide on which
models or theories are the best. This is an example proof of the
feasibility of the COMPETE program, and of its utility.

We have given here the best possible estimates
for present and future $pp$ and $\bar p p$ facilities.
The central values of our fits and the corresponding statistical error
give our ``best guess'' estimate. The systematic experimental errors
tell us how much this guess could be affected by incompatible data.
The theoretical systematic errors will tell us directly whether an
experiment can be fitted by one of the standard analytic parametrisations, or
whether it calls for new ideas.

\section*{Acknowledgments }
The COMPAS group was supported in part by the Russian Foundation for
Basic Research grants RFBR-98-07-90381 and RFBR-01-07-90392, K.K.
is in part supported by the U.S. D.o.E. Contract DE-FG-02-91ER40688-Task~A,
E.M. is a visiting fellow of the Belgian FNRS.
We thank Professor Jean-Eudes Augustin for the hospitality at
LPNHE-Universit\'e Paris 6, where part of this work was done.

\end{document}